\documentclass[useAMS,usenatbib]{mn2e}
\usepackage{multirow}
\usepackage{epsfig}
\usepackage{subfigure}
\usepackage[T1, hyphens]{url}
\usepackage{tikz}
\usepackage{pifont}
\usepackage{wasysym}
\usepackage{amssymb}
\usepackage{multicol}
\usepackage{lscape}

\makeatletter
\let\jnl@style=\rm
\def\ref@jnl#1{{\jnl@style#1}}

\def\aj{\ref@jnl{AJ}}                   
\def\araa{\ref@jnl{ARA\&A}}             
\def\apj{\ref@jnl{ApJ}}                 
\def\apjl{\ref@jnl{ApJ}}                
\def\apjs{\ref@jnl{ApJS}}               
\def\ao{\ref@jnl{Appl.~Opt.}}           
\def\apss{\ref@jnl{Ap\&SS}}             
\def\aap{\ref@jnl{A\&A}}                
\def\aapr{\ref@jnl{A\&A~Rev.}}          
\def\aaps{\ref@jnl{A\&AS}}              
\def\azh{\ref@jnl{AZh}}                 
\def\baas{\ref@jnl{BAAS}}               
\def\jrasc{\ref@jnl{JRASC}}             
\def\memras{\ref@jnl{MmRAS}}            
\def\mnras{\ref@jnl{MNRAS}}             
\def\pra{\ref@jnl{Phys.~Rev.~A}}        
\def\prb{\ref@jnl{Phys.~Rev.~B}}        
\def\prc{\ref@jnl{Phys.~Rev.~C}}        
\def\prd{\ref@jnl{Phys.~Rev.~D}}        
\def\pre{\ref@jnl{Phys.~Rev.~E}}        
\def\prl{\ref@jnl{Phys.~Rev.~Lett.}}    
\def\pasp{\ref@jnl{PASP}}               
\def\pasj{\ref@jnl{PASJ}}               
\def\rmxaa{\ref@jnl{RMXAA}}             
\def\qjras{\ref@jnl{QJRAS}}             
\def\skytel{\ref@jnl{S\&T}}             
\def\solphys{\ref@jnl{Sol.~Phys.}}      
\def\sovast{\ref@jnl{Soviet~Ast.}}      
\def\ssr{\ref@jnl{Space~Sci.~Rev.}}     
\def\zap{\ref@jnl{ZAp}}                 
\def\nat{\ref@jnl{Nature}}              
\def\iaucirc{\ref@jnl{IAU~Circ.}}       
\def\aplett{\ref@jnl{Astrophys.~Lett.}} 
\def\apspr{\ref@jnl{Astrophys.~Space~Phys.~Res.}}
\def\bain{\ref@jnl{Bull.~Astron.~Inst.~Netherlands}}
\def\fcp{\ref@jnl{Fund.~Cosmic~Phys.}}  
\def\gca{\ref@jnl{Geochim.~Cosmochim.~Acta}}   
\def\grl{\ref@jnl{Geophys.~Res.~Lett.}} 
\def\jcp{\ref@jnl{J.~Chem.~Phys.}}      
\def\jgr{\ref@jnl{J.~Geophys.~Res.}}    
\def\jqsrt{\ref@jnl{J.~Quant.~Spec.~Radiat.~Transf.}}
\def\memsai{\ref@jnl{Mem.~Soc.~Astron.~Italiana}}
\def\nphysa{\ref@jnl{Nucl.~Phys.~A}}   
\def\physrep{\ref@jnl{Phys.~Rep.}}   
\def\physscr{\ref@jnl{Phys.~Scr}}   
\def\planss{\ref@jnl{Planet.~Space~Sci.}}   
\def\procspie{\ref@jnl{Proc.~SPIE}}   

\makeatother

\newcommand {\apgt} {\ {\raise-.5ex\hbox{$\buildrel>\over\sim$}}\ }
\newcommand {\aplt} {\ {\raise-.5ex\hbox{$\buildrel<\over\sim$}}\ } 

\title[The XMM and NuSTAR view of NGC 2992]{Tracking the Iron K$\alpha$ line and the Ultra Fast Outflow in NGC 2992 at different accretion states}
\author[Andrea Marinucci, et al.]{A. Marinucci$^{1}$\thanks{E-mail: marinucci@fis.uniroma3.it (AM)}, S. Bianchi$^{1}$, V. Braito$^{2,3}$, G. Matt$^{1}$, E. Nardini$^{4}$, J. Reeves$^{2,5}$\\
\\
$^1$Dipartimento di Matematica e Fisica, Universit\`a degli Studi Roma Tre, via della Vasca Navale 84, 00146 Roma, Italy\\
$^2$Center for Space Science and Technology, University of Maryland Baltimore County, 1000 Hilltop Circle, Baltimore, MD 21250, USA\\
$^3$INAF-Osservatorio Astronomico di Brera, Via Bianchi 46 I-23807 Merate (LC), Italy\\
$^4$Istituto Nazionale di Astrofisica (INAF) – Osservatorio Astrofisico di Arcetri, Largo Enrico Fermi 5, 50125 Firenze, Italy\\
$^5$Astrophysics Group, School of Physical and Geographical Sciences, Keele University, Keele, Staffordshire, ST5 5BG, UK\\
}
\begin{document}
\maketitle
\label{firstpage}
\begin{abstract}  
The Seyfert 2 galaxy NGC 2992 has been monitored eight times by XMM-{\it Newton} in 2010 and then observed again in 2013, while in 2015 it was simultaneously targeted by {\it Swift} and {\it NuSTAR}. XMM-{\it Newton} always caught the source in a faint state (2-10 keV fluxes ranging from 0.3 to 1.6$\times10^{-11}$ erg cm$^{-2}$ s$^{-1}$) but {\it NuSTAR} showed an increase in the 2-10  keV flux up to 6$\times10^{-11}$ erg cm$^{-2}$ s$^{-1}$.  We find possible evidence of an Ultra Fast Outflow with velocity $v_1=0.21\pm0.01c$ (detected at about 99\% confidence level) in such a flux state.
The UFO in NGC 2992 is consistent with being ejected at a few tens of gravitational radii only at accretion rates greater than 2\% of the Eddington luminosity. 
The analysis of the low flux 2010/2013 XMM data allowed us to determine that the Iron K$\alpha$ emission line complex in this object is likely the sum of three distinct components: a constant, narrow one due to reflection from cold, distant material (likely the molecular torus); 
a narrow, but variable one which is more intense in brighter observations and a broad relativistic one emitted in the innermost regions of the accretion disk, which has been detected only in the 2003 XMM observation.
\end{abstract}

\section{Introduction}
\begin{figure*}
 \epsfig{file=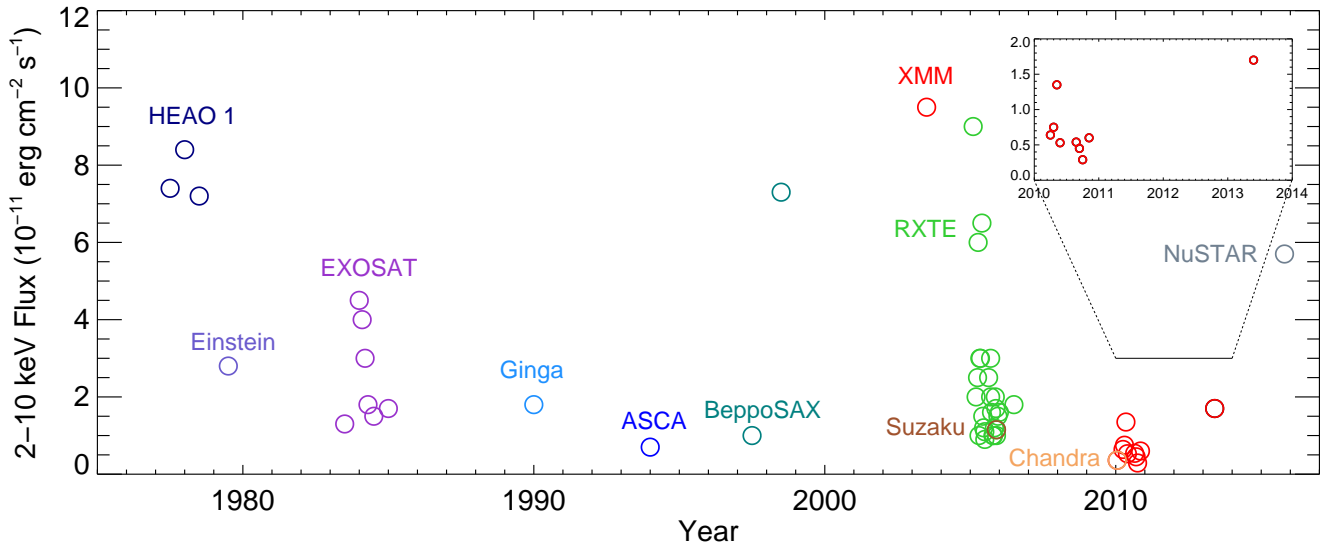, width=2.14\columnwidth}\\
  \caption{Historical 2-10 keV light curve of the source. This plot is an extension of the one previously presented in \citet{mkt07}. The 2010/2013  XMM fluxes are shown in red in the top-right inset.}
  \label{lcurve}
\end{figure*}
Variability is one of the best tools to investigate the emission mechanisms in Active Galactic Nuclei (AGN). While in many cases flux variations can be attributed to opacity changes in the line-of-sight absorbers \citep[e.g. NGC 1365 and NGC 1068:][]{ris05,mbm16}, a few sources have been observed to vary dramatically in the X-ray intrinsic flux. Highly variable AGN are the perfect astrophysical laboratories for studying the accretion/ejection mechanisms occurring at different radii in the accretion disk, in response to the primary continuum variations (i.e. at different accretion rates). The  profile of the Iron K$\alpha$ emission line gives information about the relativistic effects occurring in the innermost regions, about the emissivity of the disk and its ionization state \citep{frs89,mama96,fab00,rn03,wf12,rey14}. On the other hand, accretion disk winds ejected at larger radii are investigated by studying the blueshifted absorption features above $\sim$7 keV \citep{tcr10,tcr11, tcr12} and they give us indications on the energetic feedback to the host galaxy (King \& Pounds 2015).  Long monitorings in the X-rays (using in particular the XMM spectral resolution combined with the {\it NuSTAR} broad energy coverage) have provided a number of results in the framework of black hole spin measurements (MCG-6-30-15: Marinucci et al. 2014a; NGC 1365: Walton et al. 2014; SWIFT J2127.4: Marinucci et al. 2014b, Mrk 335: Parker et al., 2014) and Ultra Fast Outflows measurements (PDS 456: Nardini et al. 2015; IRAS 13224-3809: Parker et al. 2017).  

NGC 2992 is a highly inclined (i=70$^{\circ}$) spiral galaxy at z=0.00771 \citep{keel96}, classified as a Seyfert 1.9/1.5 \citep{trippe08}. In the X-rays, it is absorbed by a column density $N_{\rm H} \sim9\times10^{21}$ cm$^{-2}$ and it steadily declined in observed flux from 1978, when it was observed by HEAO1 (Mushotzky 1982) at a flux level of about $8\times10^{-11}$ erg cm$^{-2}$ s$^{-1}$ until 1994, when it was observed by {\it ASCA} \citep{wny96} at a flux level by about a factor 20 fainter (see Fig. 1). Then it underwent a rapid recovery: in 1997 it was observed by BeppoSAX at a flux level somewhat higher than in 1994, while in 1998 the source fully recovered its HEAO-1 brightness \citep{gilli00}. In 2003, when observed by XMM-{\it Newton}, the flux was even higher, about 10$^{-10}$ erg cm$^{-2}$ s$^{-1}$, but the observation was in Full Frame mode, and so inevitably piled-up. Nevertheless, an intense relativistic component of the iron K$\alpha$ emission line was claimed (EW=$250\pm70$ eV: Shu et al. 2010). The source was then observed by {\it Suzaku} on November/December 2005, and found in a much fainter state, almost an order of magnitude fainter than in the XMM-{\it Newton} observation. Both an unresolved and a broad component of the iron line were detected (Yaqoob et al. 2007).
The time behavior of the iron line is very interesting, suggesting the presence of a relativistic component which becomes more extreme at high flux levels. Notably, this behavior is the opposite of what is observed in other sources with relativistic lines and explained in the framework of the light bending model \citep{mama96,mf04}.

The {\it RXTE} monitoring campaign (Fig. 1) consisted of 24 pointings between early March 2005 and late January 2006, with the interval between observations ranging from 3 to 33 days.  Large amplitude (almost an order of magnitude) variability was found, indicating that the variability timescale is quite short, of the order of days, while no significant variation of the primary power law index was found. 
Similar flux variations, up to a factor $\sim 10$, are also apparent in the Swift/BAT light curve. Two flares (with 2-10 keV fluxes  F$_{\rm x}$=$3.6\times10^{-10}$ erg cm$^{-2}$ s$^{-1}$) have been recently measured on June, 2 2016 and on April, 14 2014 by the MAXI team in a sky position consistent with NGC 2992 \citep{nut16}. In 2010, NGC 2992 was observed eight times for $\sim 40$ ks by XMM-{\it Newton} and three times by {\it Chandra} \citep{mn17},  with a 2--10 keV flux ranging from $∼3\times10^{-12}$ erg cm$^{-2}$ s$^{-1}$ (its historical minimum) to $1.3\times10^{-11}$ erg cm$^{-2}$ s$^{-1}$. A further 2013 XMM-{\it Newton} observation caught the source in a similarly low flux state ($1.6\times10^{-11}$ erg cm$^{-2}$ s$^{-1}$). The most recent X-ray observation of this object was performed simultaneously with {\it Swift} and {\it NuSTAR} in 2015 and a 2-10 keV flux of $\sim6\times10^{-11}$ erg cm$^{-2}$ s$^{-1}$ was measured. As suggested by \citet{ymt07} the overall behavior suggests a scenario in which the accretion disk becomes more radiatively efficient at high luminosities, making NGC 2992 one of the few sources in which the broad, relativistic component of the Iron K$\alpha$ is seen to respond to nuclear variations.  

We hereby analyse the latest XMM, {\it Swift} and {\it NuSTAR} spectra, with the aim of understanding the physical scenario beneath this puzzling source. The paper is structured as follows: in Sect. 2 we discuss the observations and data reduction, in Sect. 3  we present the spectral analyses. We discuss and summarize the physical implications of our results in Sect. 4 and 5. 

\begin{figure*}
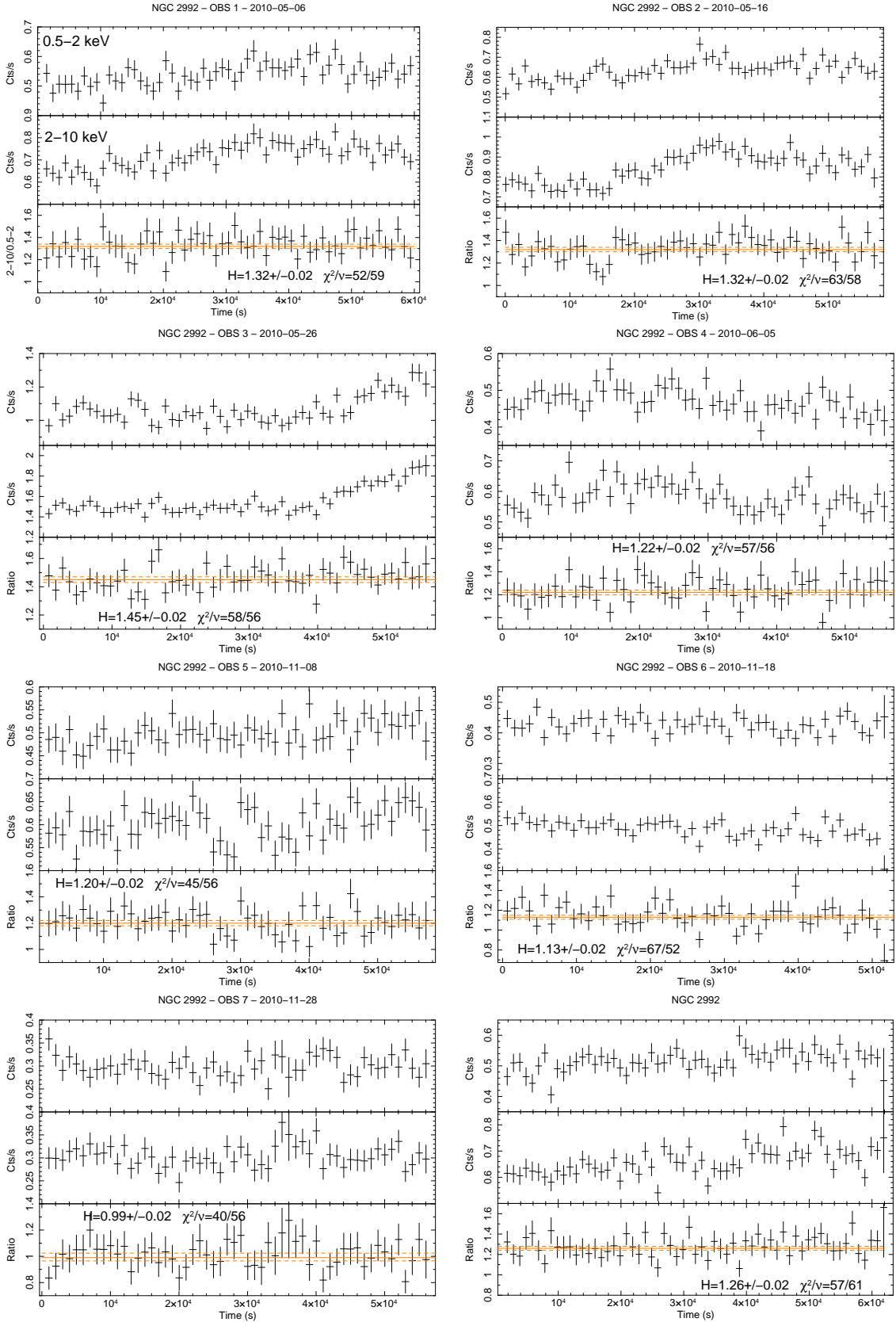

 \epsfig{file=obs1_lc.ps, angle=-90, width=0.9\columnwidth}
  \epsfig{file=obs2_lc.ps, angle=-90, width=0.9\columnwidth}
    \epsfig{file=obs3_lc.ps, angle=-90, width=0.9\columnwidth}
    \epsfig{file=obs4_lc.ps, angle=-90, width=0.9\columnwidth}
    \epsfig{file=obs5_lc.ps, angle=-90, width=0.9\columnwidth}
    \epsfig{file=obs6_lc.ps, angle=-90, width=0.9\columnwidth}
    \epsfig{file=obs7_lc.ps, angle=-90, width=0.9\columnwidth}
    \epsfig{file=obs8_lc.ps, angle=-90, width=0.9\columnwidth}
  \caption{Light curves and hardness ratios for the eight observations of the XMM monitoring campaign. We used a time binning of 1000 s.}
  \label{lcurves}
\end{figure*}

\section{Observations and data reduction}
\subsection{XMM-Newton}
NGC 2992 (z=0.0071) has been subject to a monitoring campaign with XMM-{\it Newton} \citep{xmm} in 2010,  starting on 2010 May 6 with the EPIC CCD cameras, the Pn \citep{struder01} and the two MOS \citep{turner01}, operated in small window and thin filter mode. Data from the MOS detectors are not included in our analysis due to the lower statistics of the spectra. The extraction radii and the optimal time cuts for flaring particle background were computed with SAS 16 \citep{gabr04} via an iterative process which leads to a maximization of the SNR, similar to the approach described in \citet{pico04}. The resulting optimal extraction radii, net exposure times, count rates and ObsIDs are listed in Table \ref{logobs} and the background spectra were extracted from source-free circular regions with a radius of 50 arcsec. The 2003 observation (ObsID. 0147920301) is heavily affected by pile-up and we therefore considered a source extraction annulus from 10 to 40 arcsec to remove this effect. After careful inspection of the { \it epatplot} output, confirming the goodness of the data, the net exposure time is 23 ks. This is in agreement with the previous analysis of this data set presented in \citet{sym10}.\\
Spectra were then binned in order to over-sample the instrumental resolution by at least a factor of three and to have no less than 30 counts in each background-subtracted spectral channel. Light curves in the 0.5-2 keV, 0.5-10 keV bands and hardness ratios can be seen in Fig. \ref{lcurves}: since no significant spectral variability is observed within each observation we used time averaged spectra.
We adopt the cosmological parameters $H_0=70$ km s$^{-1}$ Mpc$^{-1}$, $\Omega_\Lambda=0.73$ and $\Omega_m=0.27$, i.e. the default ones in \textsc{xspec 12.9.0} \citep{xspec}. Errors correspond to the 90\% confidence level for one interesting parameter ($\Delta\chi^2=2.7$), if not stated otherwise. 

\begin{table}
\begin{center}
\begin{tabular}{ccccc}
{\bf Obs. ID} & {\bf Date} & {\bf Time} & {\bf Count rate}& {\bf Region} \\
 &  & {\bf  (ks)} & {\bf  (cts/s)} & {\bf (arcsec)}\\
\hline
0654910301& 2010-05-06    &  41.1& $1.304\pm 0.170$ &25 \\
 0654910401 & 2010-05-16   &     40.2&  $1.526\pm 0.049$& 39 \\
 0654910501  &   2010-05-26  &    38.5 &  $2.688\pm0.018$& 35\\
  0654910601& 2010-06-05   & 38.7 & $1.094\pm0.096$& 28\\
0654910701  & 2010-11-08 & 38.8& $1.146\pm0.041$&40 \\
0654910801  &  2010-11-18  & 35.8&$0.953\pm0.037$ & 36\\
0654910901  & 2010-11-28  &37.5 & $0.641\pm0.076$& 26\\
 0654911001 &  2010-12-08  &42.0 & $1.218\pm 0.061$&37 \\
0701780101  & 2013-05-11  & 9.0& $3.301\pm 0.047$& 40\\
\hline
60160371002 &    2015-12-02& 20.8& $2.411\pm0.047$& 50\\
00081055001 & 2015-12-02& 6.5& $0.809\pm0.010$& 120\\
\hline
\end{tabular}
\end{center}
\caption{\label{logobs} Observation log for the NGC 2992 observations with XMM-{\it Newton} (2010--2013), {\it NuSTAR} and {\it Swift} (2015). XMM and {\it Swift} count rates are inferred in the 0.5-10 keV energy band, while for {\it NuSTAR} we used 3-79 keV.}
\end{table}
\subsection{NuSTAR}
{\it NuSTAR} (Harrison et al. 2013) observed NGC 2992 with its two coaligned X-ray telescopes with corresponding Focal Plane Module A (FPMA) and B (FPMB) on 2015 December 2  for a total of $\sim37.4$ ks of elapsed time, respectively.  The Level 1 data products were processed with the {\it NuSTAR} Data Analysis Software (NuSTARDAS) package (v. 1.7.1). Cleaned event files (level 2 data products) were produced and calibrated using standard filtering criteria with the \textsc{nupipeline} task and the latest calibration files available in the {\it NuSTAR} calibration database (CALDB 20170720). Extraction radii for the source and background spectra were $50$ arcsec and $70$ arcsec, respectively. After this process, the net exposure times for the two observations were 20.8 ks. Count rates, ObsID. and net exposure time are summarized in Table \ref{logobs}. The two {\it NuSTAR} spectra were binned in order to over-sample the instrumental resolution by at least a factor of 2.5 and to have a Signal-to-Noise Ratio (SNR) greater than 5$\sigma$ in each spectral channel. A cross-calibration factor $K_1=1.017\pm0.013$ between the two detectors is found.

\subsection{Swift}
As part of the {\it Swift}-BAT AGN survey, the source has been also observed by {\it Swift} XRT on 2015 December 2, simultaneously to {\it NuSTAR}. Using a 120 arcsec radius circular region centered on the source we find a count rate of $0.81\pm0.01$ cts/s in the whole energy band, which is greater than the $\sim$0.6 cts/s threshold for pile-up \citep{moretti04, vgb06}. Following the online guideline\footnote{\url{http://www.swift.ac.uk/analysis/xrt/pileup.php}} we excluded the central 8 arcsec and used a background region of 120 arcsec radius. The resulting spectrum was binned in order to over-sample the instrumental resolution by at least a factor of 3 and to have at least 30 counts in each spectral channel. We allowed for a cross-calibration constant $K_2=0.97\pm0.05$ between XRT and FPMA spectra.
\begin{figure}
 \epsfig{file=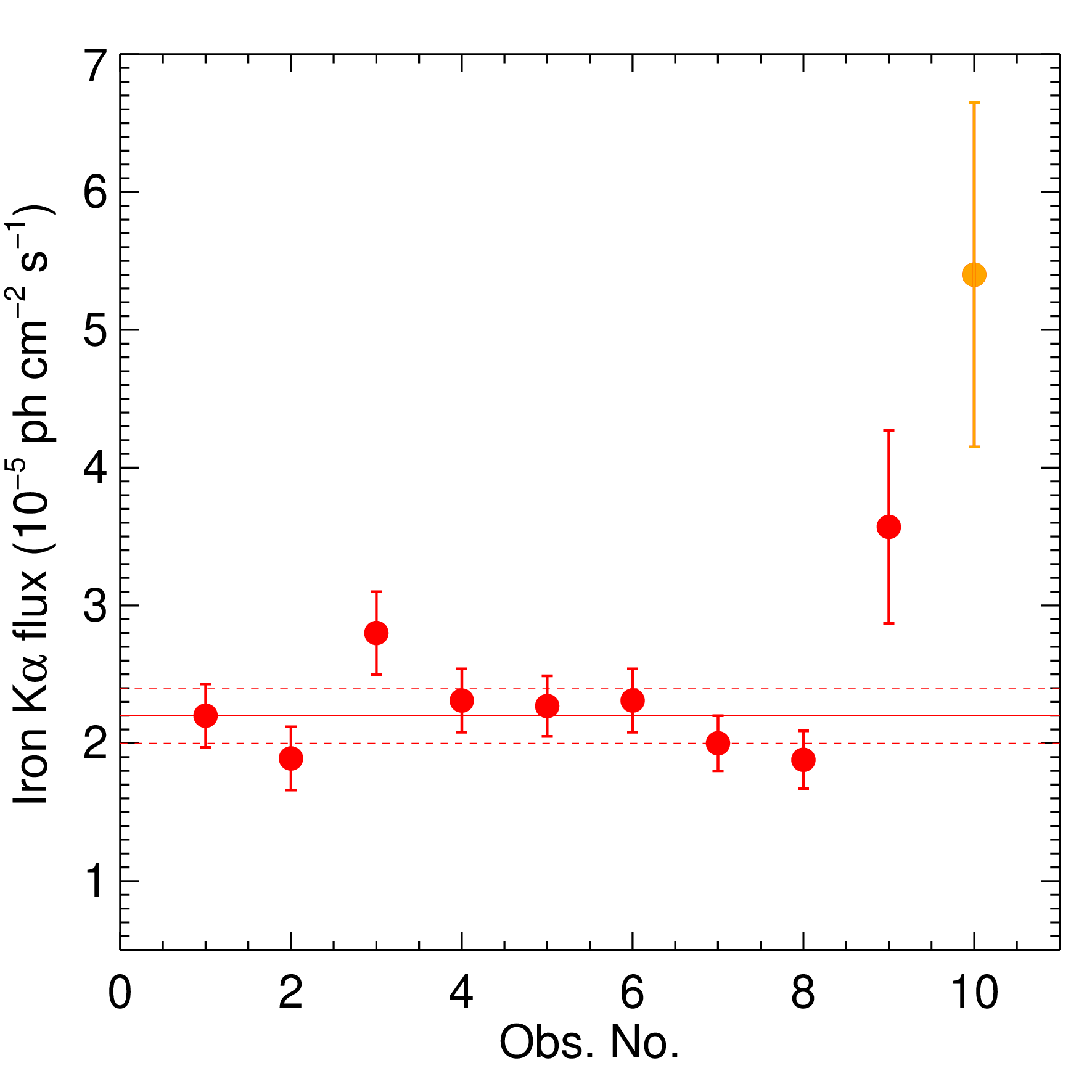, width=0.49\columnwidth}
  \epsfig{file=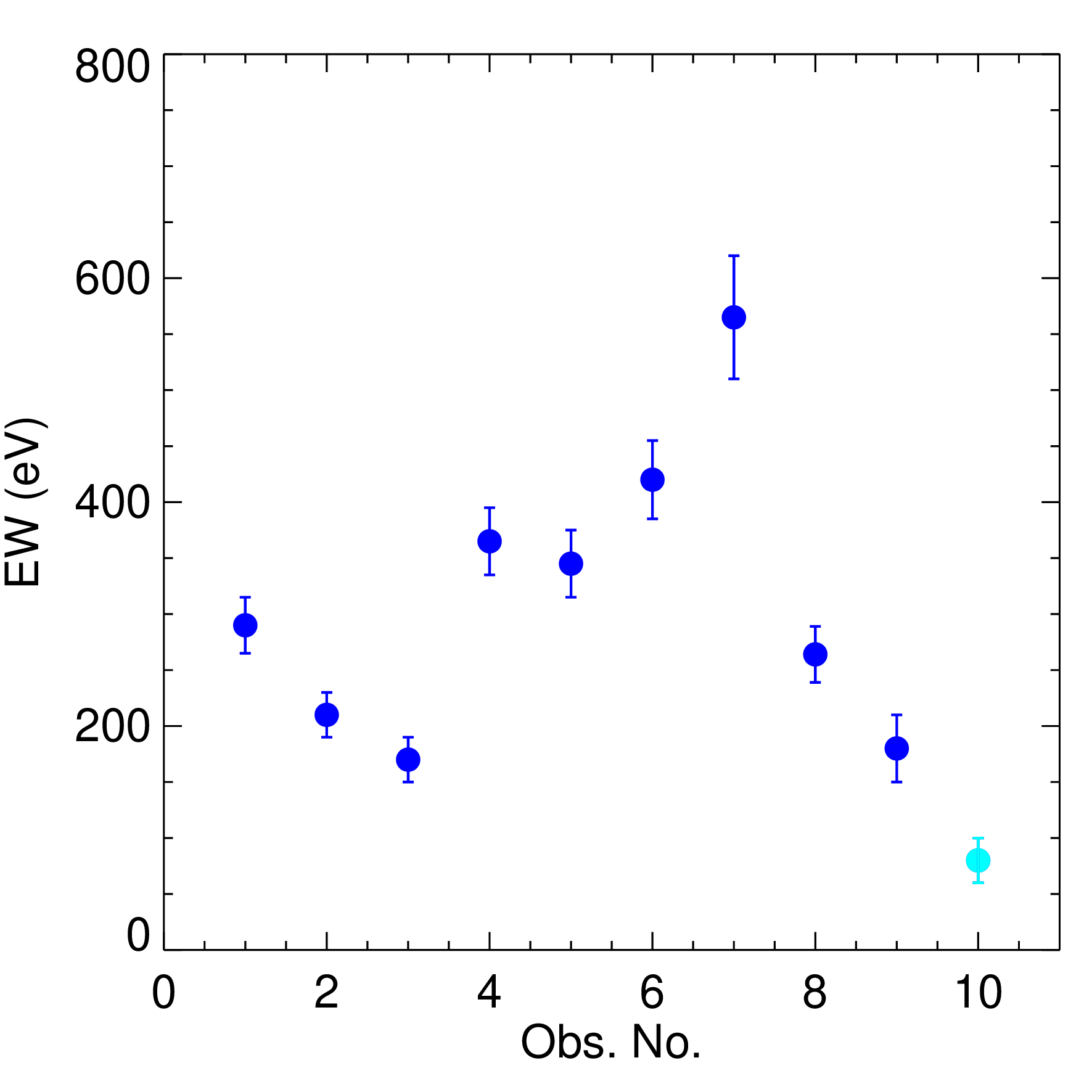, width=0.49\columnwidth}
  \caption{Temporal behavior of the Fe K$\alpha$ flux and EW (when only a single, narrow component of the line is included in the model). Horizontal solid and dashed red lines indicate best fit value and error bars when the Fe K$\alpha$ flux is tied between the nine spectra. Red and blue data points are XMM pointings, orange and cyan refer to Swift-{\it NuSTAR} data. We adopted 1$\sigma$ error bars.}  \label{510}
\end{figure}
\begin{figure*}
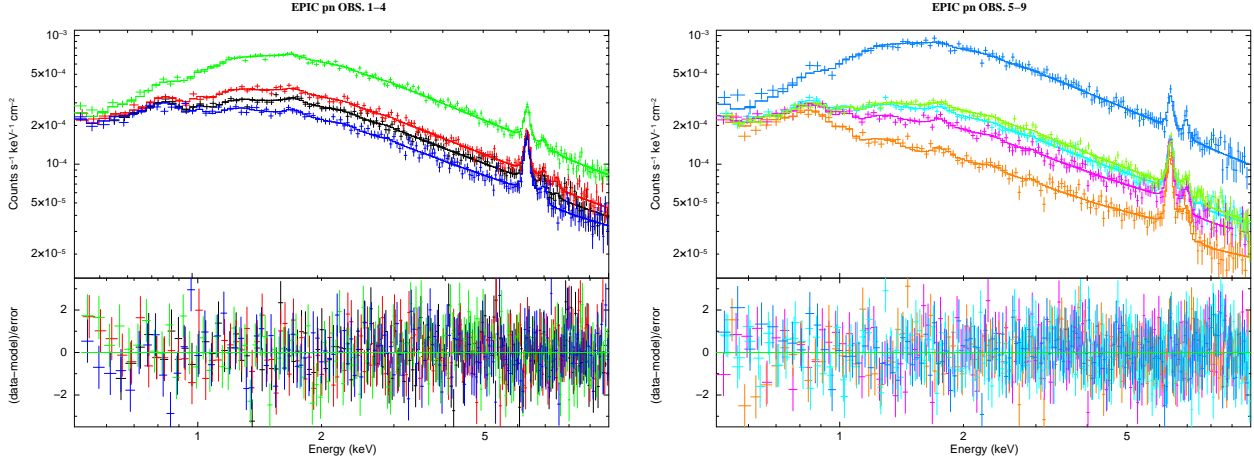

 \epsfig{file=xmm14.ps, angle=-90, width=\columnwidth}
  \epsfig{file=xmm59.ps, angle=-90, width=\columnwidth}
  \caption{EPIC pn best fits, with associated residuals. All spectra are divided by the effective area of the instrument, for plotting purposes only.}
  \label{xmm_best}
\end{figure*}
\begin{table*}
\begin{center}
\begin{tabular}{cccccccc}
{\bf Obs. Date} & \multicolumn{7}{c}{\bf Best fit parameter} \\
\hline
 & N$_{\rm H}$ & $\Gamma$ & N$_{\rm pow}$ & N$_{\rm xill}$ &$R$ &$F_{\rm 2-10\ keV}$ & $L_{\rm 2-10\ keV}$ \\
 & $(\times10^{22}$ cm$^{-2})$& &$(\times10^{-3})$ & $(\times10^{-4})$&  & $(\times10^{-11}$ erg cm$^{-2}$ s$^{-1})$ &   $(\times10^{42}$ erg s$^{-1})$ \\
\hline
 2010-05-06& $0.92\pm0.06$&$1.67\pm0.05$ &$1.30\pm0.05$ & $1.04\pm0.08$& $1.65\pm0.20$&$0.65\pm0.03$&$0.90\pm0.05$\\
 2010-05-16& $0.86\pm0.05$& $1.64\pm0.04$& $1.50\pm0.07$& $1.01\pm0.08$& $1.25\pm0.15$&$0.75\pm0.02$ & $1.05\pm0.05$\\
 2010-05-26&$0.84\pm0.04$ & $1.61\pm0.04$&$2.78\pm0.08$ & $1.45\pm0.15$& $0.90\pm0.15$&$1.35\pm0.02$& $1.90\pm0.10$\\
 2010-06-05&$0.85\pm0.08$ & $1.67\pm0.05$&$1.00\pm0.05$ & $1.07\pm0.07$& $2.25\pm0.35$&$0.53\pm0.03$ & $0.75\pm0.05$\\
 2010-11-08& $0.80\pm0.07$& $1.67\pm0.05$&$1.03\pm0.05$ & $1.10\pm0.07$& $2.20\pm0.25$&$0.54\pm0.03$ &  $0.75\pm0.04$\\
 2010-11-18&$0.80\pm0.08$ &$1.70\pm0.06$ &$0.80\pm0.05$ & $1.10\pm0.07$& $3.10\pm0.50$&$0.45\pm0.02$ &  $0.65\pm0.05$\\
 2010-11-28&$0.81\pm0.10$ &$1.71\pm0.09$ & $0.45\pm0.04$& $0.85\pm0.10$&$4.50\pm1.10$ &$0.30\pm0.02$ &$0.40\pm0.03$ \\
 2010-12-08&$0.90\pm0.06$ &$1.68\pm0.04$ &$1.20\pm0.06$ & $0.95\pm0.07$& $1.65\pm0.20$&$0.60\pm0.02$ &  $0.85\pm0.03$\\
 2013-05-11& $0.81\pm0.05$& $1.63\pm0.06$&$3.45\pm0.05$ & $1.78\pm0.28$& $0.96\pm0.20$&$1.65\pm0.05$& $2.30\pm0.10$\\
\hline
\end{tabular}
\end{center}
\caption{\label{bestfitPar} Best fit parameters of the combined XMM analysis. Luminosities are corrected for absorption.}
\end{table*}
\section{Spectral analysis}
\subsection{The 2010-2013 XMM-Newton low flux states}
We start our spectral analysis of the 2010/2013 XMM data by fitting the 5-10 keV spectra with the aim of searching for variations of the iron K$\alpha$ line.  As a first step, we only leave the continuum free to vary and lines parameters tied. We used an absorbed power law and 5 Gaussians to model the following emission lines: neutral Fe K$\alpha$ and K$\beta$, Fe \textsc{xxv} He-$\alpha$,  Fe \textsc{xxvi} Ly-$\alpha$ and Ni K$\alpha$. We also included a \textsc{zashift} component in \textsc{xspec} to take into account the mis-calibration of the EPIC-pn CTI \citep[see][for Ark 120, NGC 5548 and MCG-6-30-15, respectively]{npr16,cdp16,mmm14}.  At this stage, the only variable parameters are the power law slope and normalization, the absorbing column density and the shift of the lines. We obtain a good $\chi^2$/dof=556/516=1.07. To search for variations between the nine observations we allow fluxes and energy centroids of the emission lines free to adjust: a $\chi^2$/dof=516/507=1.01 is retrieved. Fluxes and EWs are plotted in Fig. \ref{510}.  Best fit values for emission lines other than the neutral Fe K$\alpha$ will be further discussed at the end of this Section. Fig. \ref{510}, left panel shows that the flux of the narrow Fe K$\alpha$ is constant among the nine observations, with the exception of Obs 3 and 9, and the best fit value ($2.2\pm0.2\times10^{-5}$ ph cm$^{-2}$ s$^{-1}$) is in perfect agreement with the one found with {\it Chandra} HETG \citep{mn17}. The change of the Iron K$\alpha$ flux in Obs. 3 and 9 suggests either a variation of the Compton-thick reflector or the appearance of a new spectral component. Since a further, broad iron K$\alpha$ component, arising from the innermost regions of the accretion disk was found in past {\it Suzaku} and XMM data \citep{ymt07,sym10}, we tried to include it in our fits, fixing the narrow iron K$\alpha$ to the combined best fit value. The additional 6.4 keV lines are unresolved: only upper limits for the widths $\sigma_3<$70 eV and $\sigma_9<$90 eV for Obs. 3 and 9 are found, respectively. The corresponding fluxes are $F_3= 0.6\pm0.3\times10^{-5}$ and $F_9= 1.4\pm0.6\times10^{-5}$ ph cm$^{-2}$ s$^{-1}$. These width of the lines are converted into velocities $v_3<7700$ km/s and $v_9<9900$ km/s (FWHM), suggesting that the additional component could be due to material located in the Broad Line Region but not in the inner accretion disk. For completeness, we also included in the plots the best fit parameter of the same model applied to 2014 {\it NuSTAR} data. 

We then extended the analysis down to 0.5 keV modifying the model as follows. We included an absorbed primary component ({\sc zwabs*cutoffpl} in {\sc xspec}) and cold reflection from distant material \citep[\textsc{xillver:}][]{garcia13}. We linked the photon index and high-energy cutoff of {\sc xillver} to the ones of the {\sc cutoffpl} component, fixing the inclination angle to 30 degrees, only the reflected spectrum is taken into account. Throughout the paper we will use the reflection fraction (i.e. the fraction of the illuminating continuum which is Compton scattered) as the ratio between the 10-50 keV luminosities: ${R\rm=L_{refl}/L_{int}}$. Due to the limited XMM energy band the high energy cutoff is fixed to 500 keV. A soft X-ray scattered component below $\sim2$ keV from gas photoionised by the nuclear continuum, possibly associated to the NLR \citep{bianchi06,gb07} has been included. We produced a grid model for \textsc{xspec} using \textsc{cloudy} 17 \citep[last described by][]{fcg17}. It is an extension of the same model used in \citet{bianchi10, mbm11, mbf17}. Grid parameters are $\log U=[-2.00:4.00]$, step 0.25, $\log N_\mathrm{H}=[19.0:23.5]$, step 0.1. Only the reflected spectrum, arising from the illuminated face of the cloud, has been taken into account in our model. The whole model has been then multiplied by a Galactic absorption of $4.8 \times 10^{20}$cm$^{-2}$ \citep{kalberla05}. We also included an {\sc apec} model to reproduce the thermal emission from extra-nuclear material previously observed with {\it Chandra} at low energies \citep{csv05}. At this stage of the spectral analysis, the photon index is initially fixed to $\Gamma=1.7$, the normalization of the Compton reflection ({\sc xillver}) is tied between the nine observations and the fluxes of the emission lines in the 6.5-7.5 keV interval are free. Due to the variable intrinsic flux of the source, we also left the absorbing column density and power law normalization free to vary. The extra-nuclear, soft emission parameters ($\log U$, $\log N_\mathrm{H}$, normalization in {\sc cloudy} and kT, normalization in {\sc apec}) are tied between the nine observations. The reduced $\chi^2$ is decent ($\chi^2$/dof=1727/1351=1.28) but some residuals can be found around $\sim$1 keV and above 5 keV.  The inclusion of a narrow Gaussian improves the fit ($\Delta\chi^2$=-76 for 2 additional degrees of freedom) with a resulting energy E$=1.05\pm0.03$ keV and flux F$=0.5\pm0.1\times10^{-5}$ ph cm$^{-2}$ s$^{-1}$ and consistent with a blend of Ne {\sc x} K$\alpha$ and Fe {\sc xvii} 4C emission lines. Detailed modeling of the soft X-ray emission lines is beyond the purpose of this paper and will be addressed, taking into account the RGS spectra, in a forthcoming paper. We then left the {\sc xillver} normalization and the photon index free to change between the observations, with statistical improvements $\Delta\chi^2$=-175/8 d.o.f. and $\Delta\chi^2$=-92/9 d.o.f., respectively. We obtain a  best fit  $\chi^2$/dof=1384/1332=1.03. Best fits, spectra and residuals can be seen in Fig. \ref{xmm_best} and Table \ref{bestfitPar}. Best fit parameters for the soft X-ray emission are $\log U$=$1.68^{+0.10}_{-0.14}$, $\log$(N$_{\rm H}/{\rm cm^{-2}})$=$21.1^{+0.2}_{-0.3}$ and kT$=0.45\pm0.05$, for the {\sc cloudy} and {\sc apec} components, respectively. Their 0.5-2.0 keV observed luminosities are L$_{\rm CL}$=$1.8^{+1.3}_{-0.7}\times10^{40}$ erg/s and L$_{\rm AP}$=$6.8^{+1.0}_{-0.6}\times10^{39}$ erg/s.

\begin{figure}
 \epsfig{file=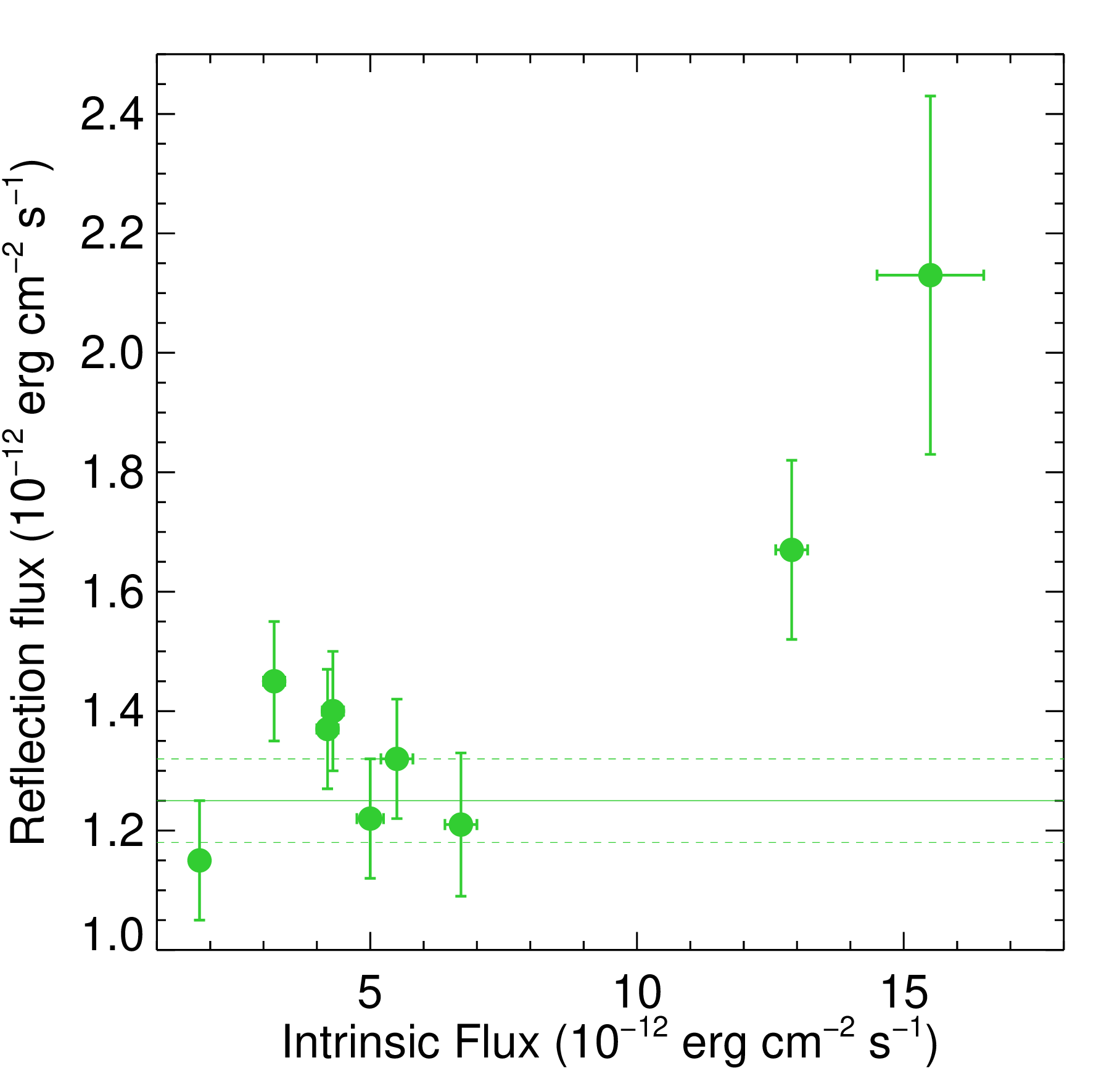, width=0.49\columnwidth}
 \epsfig{file=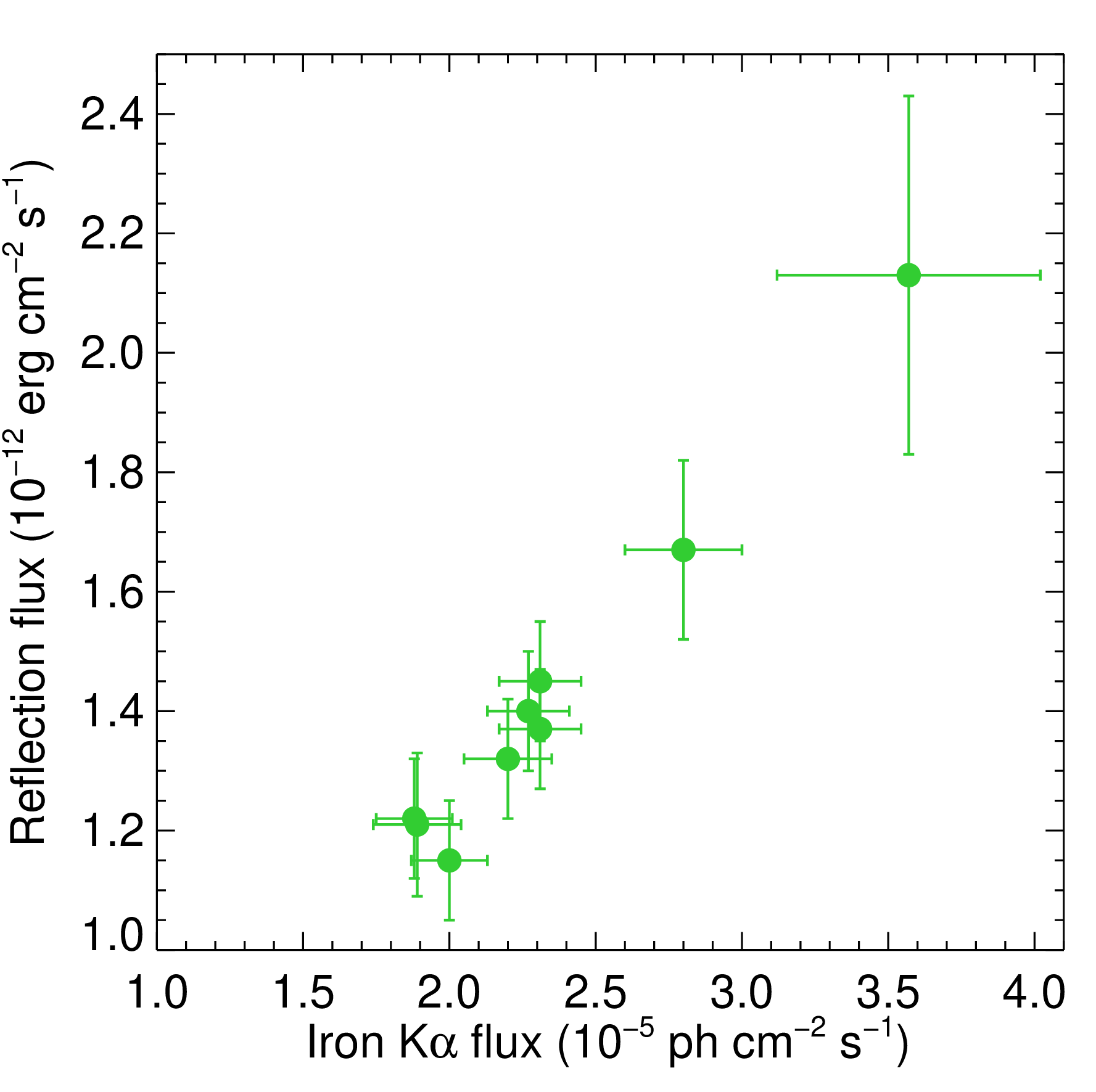, width=0.49\columnwidth}
  \caption{\label{Rfluxes} {\it Left panel:} 2-10 keV fluxes for the reflection and primary components are shown. Solid and dashed lines indicate best fit value and error bars when data are fitted simultaneously, excluding observation 3 and 9 (i.e. the ones with intrinsic fluxes greater than 1$\times10^{-11}$ erg cm$^{-2}$ s$^{-1}$). {\it Right panel:} 2-10 keV fluxes for the reflection component is plotted against the flux of the Iron K$\alpha$ emission line.}
\end{figure}
\begin{figure}
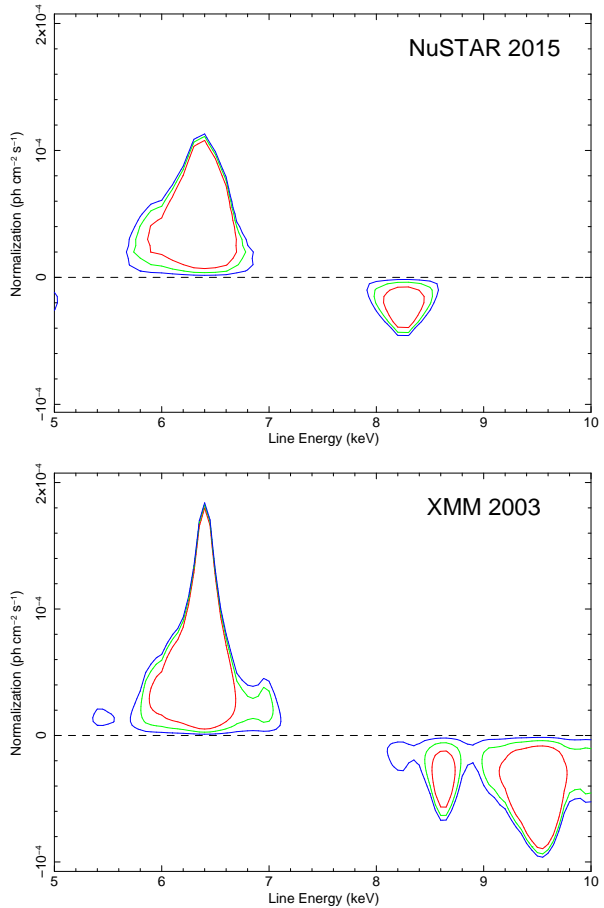

 \epsfig{file=cont_NuSTAR.ps, angle=-90, width=\columnwidth}
 \epsfig{file=cont_XMM.ps, angle=-90, width=\columnwidth} 
  \caption{Contour plots at 68\%, 90\% and 99\% confidence levels ($\Delta\chi^2$ = 2.3, 4.61, and 9.21, respectively) between the normalization and line energy when a Gaussian is left free to vary in the 5-10 keV energy range, for the 2015 {\it NuSTAR} (FPMA and FPMB, top panel) and 2003 XMM high flux observations (bottom panel). The adopted model for the continuum is composed of an absorbed power law.}
  \label{bf1}
\end{figure}

The amount of Compton reflection (i.e. the 2-10 keV flux of the {\sc xillver} component) is plotted against the intrinsic nuclear flux in Fig. \ref{Rfluxes} and solid and dashed horizontal lines indicate best fit value and error bars when all the low flux data are simultaneously fitted ($1.25\pm0.08\times10^{-12}$ erg cm$^{-2}$ s$^{-1}$). Deviations from this value are found for two observations only (Obs. 3 and 9, see also Table \ref{bestfitPar}) and this might be indicative of a response from the circumnuclear material to the primary continuum variations. While this solution is unlikely for Obs. 3 (2010-05-26) due to the short time scales between the previous and following observations (two weeks) it could be viable for the 2013 observation. We note, however, that some residuals are still found around the Iron K$\alpha$ energy and when an additional component is added ($\Delta\chi^2$=-10 for 1 additional degree of freedom) the 2-10 keV {\sc xillver} flux is consistent with being constant. We show, in the right panel of Fig. \ref{Rfluxes}, the 2-10 keV flux of the {\sc xillver} component against the total flux of the Iron K$\alpha$ emission line, from the previous analysis. The linear correlation between the two parameters clearly suggests that the {\sc xillver} component is trying to compensate for the variations of a different Fe K component \citep[similar to the Iron line behavior observed in the recent XMM campaign of Ark 120:][]{npr16}.  We conclude that the reflected emission arising from cold, distant material is constant throughout the nine observations. \\
Best fit values for energy centroids of Fe \textsc{xxv} He-$\alpha$,  Fe \textsc{xxvi} Ly-$\alpha$ and Ni K$\alpha$ (which are not accounted for by {\sc xillver}) are E$_1=6.77\pm0.09$, E$_2=7.00\pm0.05$ and E$_3=7.55\pm0.08$ keV, respectively. No statistically significant variations are found between the nine observations and their fluxes are always consistent with the ones found in the lowest state spectrum (observation 7): F$_1=0.22\pm0.09\times10^{-5}$, F$_2=0.23\pm0.11\times10^{-5}$ and F$_3=0.25\pm0.13\times10^{-5}$ ph cm$^{-2}$ s$^{-1}$.

\subsection{The 2015 Swift+NuSTAR medium flux state}
The baseline model used for the 2010 XMM spectra is then applied to the simultaneous 2015 Swift + NuSTAR observation. 
Soft X-ray emission parameters are fixed to the combined XMM fit. The resulting $\chi^2$/dof is good (418/416), and a Compton reflection fraction $R=0.18\pm0.07$ from cold material is measured, consistent with the flux of the constant XMM reflection component. Residuals around the neutral iron K$\alpha$ line and $\sim8$ keV are present (Fig. \ref{bf1}, top panel).

\begin{figure}
 \epsfig{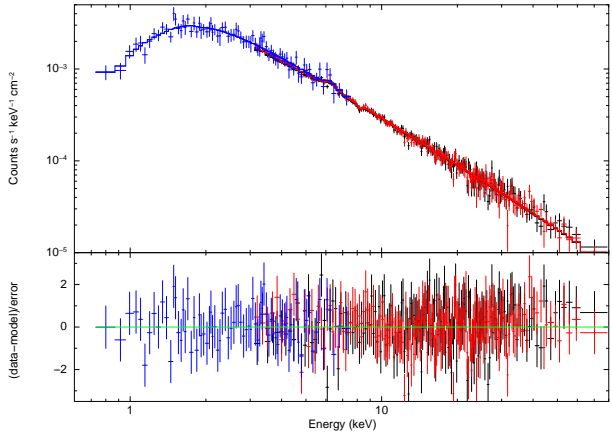}\\
  \caption{Best fit of the simultaneous Swift-{\it NuSTAR} data. Residuals are shown in the bottom panel.}
  \label{nubest}
\end{figure}
\begin{figure}
 \epsfig{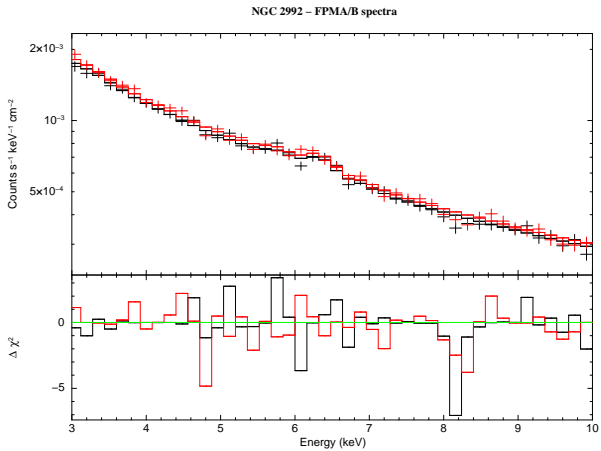}
  \caption{ The fit of the continuum model and the broad Iron K$\alpha$ applied to the data is shown, in the 3-10 keV band. We do not include {\it Swift} data for the sake of visual clarity. The 8.26 keV absorption line is clearly visible in both FPM detectors.}
  \label{UFO_res_NuSTAR}
\end{figure}

 The inclusion of two Gaussians to reproduce both emission and absorption lines leads to a best fit $\chi^2$/dof=373/411. The best fit to the data set with corresponding residuals are shown in Fig. \ref{nubest} while best fit parameters are reported in Table \ref{bestfitres}. The broad iron K$\alpha$ emission line is found at $6.32\pm0.12$ keV, with a flux F=$7.0\pm2.5\times10^{-5}$ ph cm$^{-2}$ s$^{-1}$ and a width $\sigma=250^{+190}_{-120}$ eV (with a statistical significance greater than 99.99\%, accordingly to the $F$-test). We show, in Fig. \ref{UFO_res_NuSTAR}, the residuals when the continuum and the broad Iron K$\alpha$ model is applied to the {\it NuSTAR} data, in the 3-10 keV interval. The absorption line is clearly visible in the bottom panel and it is found at $8.26^{+0.09}_{-0.12}$ keV (rest-frame energy), with a flux F=$-1.8\pm0.8\times10^{-5}$ ph cm$^{-2}$ s$^{-1}$ (with a statistical significance greater than 99.7\%).  The inclusion of the broad Iron K$\alpha$ component is accompanied by a drop in the Compton reflection fraction ($R<0.08$). This width corresponds to a velocity $v\simeq28_{-14}^{+22}\times10^3$ km/s and is consistent with material orbiting with Keplerian motion located in the outskirts of the accretion disk or in the inner regions of the BLR. The best fit power law index is $\Gamma=1.72\pm0.03$ and only a lower limit for the high energy cutoff $E_{\rm c}>350$ keV is found. Another solution could be an Iron K$\alpha$ line smeared by relativistic effects, as already discussed for the 2003 XMM high flux state of the source in \citet{sym10}. We therefore tried to model the broad line component in terms of relativistic reflection \citep[using the {\sc relxill} model in {\sc xspec};][]{dauser13}. The inclination of the disk is fixed to 30 degrees and standard values for the emissivity are assumed ($\epsilon(r)\propto r^{-3}$). The best fit solution requires a high value for the Iron abundance (A$_{\rm Fe}>7$) because of the modest Compton reflection associated to the line. The best fit reduced $\chi^2$ is statistically equivalent to the one inferred above $\chi^2$/dof=371/411 and the relativistic line has to be produced at a radius $r>35\ {\rm r_g}$. The upper limit for the cold, distant Compton reflection becomes $R<0.03$.

 In both scenarios, the reflection fraction is very low and the dominant spectral component is the primary power law continuum (with a $\Gamma=1.72\pm0.03$). We therefore find the same intensity and energy centroid for the UFO feature adopting the two different models.

\begin{table}
\begin{center}
\begin{tabular}{cc}
{\bf Parameter} & {\bf Best-fit value} \\
\hline
N$_{\rm H}$ $(\times10^{22}$ cm$^{-2})$ & $1.1\pm0.2$   \\
$\Gamma$ & $1.72\pm0.03$   \\
E$_{\rm c}$ (keV)  &     $>350$ \\
N$_{\rm pow}$  $(\times10^{-2})$&   $1.62\pm0.07$  \\
N$_{\rm xill}$ $(\times10^{-5})$ &  $<5.5$  \\
\hline
$F_{\rm 2-10\ keV}$  (erg cm$^{-2}$ s$^{-1}$) &    $5.8\pm0.3\times10^{-11}$ \\
$L_{\rm 2-10\ keV}$  (erg s$^{-1}$)  &  $7.6\pm0.3\times10^{42}$   \\
\hline
\end{tabular}
\end{center}
\caption{\label{bestfitres} Best fit parameters of the combined Swift-NuSTAR analysis.}
\end{table}
\begin{table*}
\begin{center}
\begin{tabular}{cllcccc}
 & Energy & Flux & EW & Significance & $v_{\rm out}/c$ & $\Delta\chi^2$/dof \\
 & (keV) & ($10^{-5}$ ph cm$^{-2}$ s$^{-1}$) & (eV)& $\sigma$  \\
 \hline 
2015 & $8.26^{+0.09}_{-0.12}$& $-1.8\pm0.8$& $-45\pm20$ & 2.6& $0.21\pm0.01$& -9/2\\ 
  & & &  \\
2003 & $8.61\pm0.05$&$-2.2\pm1.2$& $-40\pm25$ & 2.7 & $0.209\pm0.006$& -9/2  \\
     & $9.27\pm0.10$&$-2.0\pm1.3$& $-45\pm30$ & 1.9 & $0.31\pm0.01$& -8/2 \\
     & $9.57\pm0.06$&$-3.3\pm1.3$& $-70\pm25$ & 3.6 & $0.307\pm0.006$&  -26/2\\   
     & & & & \\
 & $\log$N$_{\rm H}$ & $\log U$ & $v_{\rm out}/c$ & $\dot{M}_{\rm out}$ & $\dot{E}_{\rm k}$ & $\dot{p}_{\rm out}$   \\
  & (cm$^{-2}$)& & & (g$\cdot$s$^{-1}$)& (erg$\cdot$s$^{-1}$) & (g$\cdot$cm$\cdot$s$^{-2}$)  \\
 \hline 
2015 & $22.25\pm0.25$& $2.45\pm0.25$& $0.21\pm0.01$ & $3.5\times10^{23}$& $6.9\times10^{42}$& $2.2\times10^{33}$  \\
     & & & & \\
2003 & $23.35_{-0.55}^{+1.10}$& $>3.1$&  $0.215\pm0.005$ & $3.8\times10^{24}$& $7.9\times10^{43}$& $2.5\times10^{34}$\\
     & $23.35_{-0.40}^{+0.15}$& $3.40_{-0.15}^{+0.40}$& $0.305\pm0.005$ & $2.7\times10^{24}$& $1.1\times10^{44}$& $2.4\times10^{34}$ \\
     \hline
\end{tabular}
\end{center}
\caption{\label{bestfitwind} Best fit parameters for the two absorbing layers, energies are reported in the rest-frame of the source. The statistical significance of the four absorption lines is calculated via Monte Carlo simulations (see Sect. 3.4 for details).}
\end{table*}

\subsection{The 2003 XMM-Newton high flux state}
\begin{figure}
 \epsfig{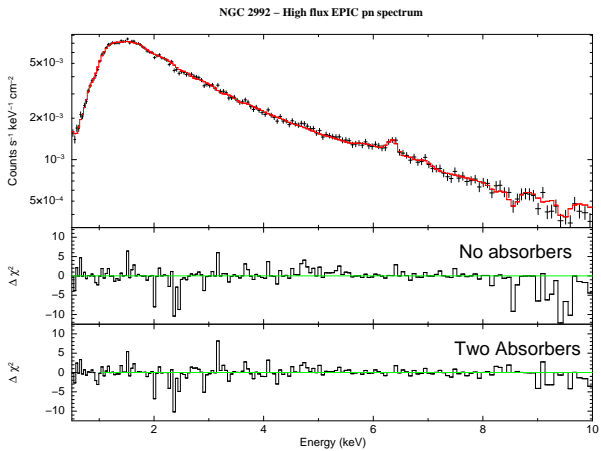}
  \caption{Best fit to the 2003, high flux, EPIC pn data and residuals are shown when a model with no absorbers (middle panel) and with the inclusion of two absorbers (bottom panel) is considered.}
  \label{xmm03}
\end{figure}

 We first apply a model composed of an absorbed primary continuum and soft X-ray, extra-nuclear emission ({\sc TBabs$\times$(zwabs$\times$cutoffpl + apec + cloudy})) to the XMM high flux state spectrum. We find a power law with $\Gamma=1.76\pm0.02$ absorbed by a column density N$_{\rm H}$=0.7$\pm0.1\times10^{22}$ cm$^{-2}$ and no statistically significant difference for the {\sc cloudy} and {\sc apec} parameters, with respect to the ones presented in Sect. 3.1. We show, in the bottom panel of Fig. \ref{bf1}, contour plots between the normalization and the energy centroid of a Gaussian line left free to vary in the 5-10 keV range. Both emission features (narrow and broad) at$\sim$6.4 keV and absorption ones above 8 keV can be clearly seen. The inclusion of a cold reflection component responsible for the narrow Iron K$\alpha$ line ($R=0.36\pm0.06$) leads to a $\chi^2$/dof=261/162=1.6 with an improvement $\Delta\chi^2=-139$ for one degree of freedom. We modeled the broad relativistic component in terms of a simple Gaussian line and our best fit values (E$_{\rm br}=6.0^{+0.2}_{-0.4}$ keV, F$_{\rm br}$=$2.4^{+0.8}_{-0.4}\times 10^{-4}$ ph cm$^{-2}$ s$^{-1}$, EW$_{\rm br}=230^{+60}_{-40}$ eV) are in perfect agreement with \citet{sym10}, leading to a $\chi^2$/dof=203/159=1.3. Residuals to this fit are shown in the middle panel of Fig. \ref{xmm03}, suggesting the presence of multiple absorption lines above 8 keV. We therefore included three narrow, unresolved Gaussians with variable energy and flux: best fit values for the rest-frame energy centroids, fluxes, EW and fit improvements are shown in Table \ref{bestfitwind}. The final goodness of fit $\chi^2$/dof=160/151=1.06 and we find a steeper continuum ($\Gamma=1.82\pm0.02$) with respect to the 2010 XMM-Newton spectra. We note, however, that a statistically comparable fit is obtained if a single broad absorption line ($\sigma=0.30^{+0.22}_{-0.15}$ keV) is adopted at 9.47$^{+0.20}_{-0.15}$ keV with an EW=-180$^{+80}_{-120}$ eV. We will discuss the physical interpretation of this component in Sect. 3.5. None of these absorption lines are present in the nine 2010/2013 XMM low flux spectra and the inclusion of an absorption line at 8.26 keV only leads to lower limits on its flux (F$_{\rm abs}>-4\times10^{-6}$ ph cm$^{-2}$ s$^{-1}$ in Obs. 3).

\subsection{Statistical significance of absorption lines}
 It is widely known that in order to assess the statistical significance of narrow unresolved features, standard likelihood ratio tests might lead to inaccurate results \citep{prot02}. We therefore followed the approach discussed in \citet{tcr10} and \citet{wmp16} to produce Monte Carlo simulations of our data sets and to retrieve a statistical significance for each absorption line detected. We produced 10000 fake data sets of the {\it NuSTAR} FPMA/B or XMM-{\it Newton} EPIC pn spectra for each of the four absorption lines reported in Table \ref{bestfitwind}, using the {\sc fakeit} command in {\sc xspec}. Responses, background files, exposure times and energy binning are the same as the ones used for real data. We considered the best fit models presented in Sect. 3.2 and 3.3 without absorption lines to simulate a fake spectrum, fitted it with the continuum model and recorded the best fitting parameters. This new model for the continuum is then used to simulate the fake data, to reduce uncertainties on the null-hypothesis probability itself \citep{pru04,mf06,mrb06}. A new unresolved Gaussian line was included in the model, and its normalization was initially set to zero and free to vary in the range $[-1.0:+1.0]\times10^{-4}$ ph cm$^{-2}$ s$^{-1}$. The energy centroid was free to vary between 7.0 and 10.0 keV in 100 eV steps and the resulting $\Delta\chi^2$ was recorded. If $N$ is the number of data sets in which a chance improvement is found and $S$ is the total number of simulated spectra, then the estimated statistical significance of the detection from Monte Carlo simulations is $1-N/S$. As an example, we can consider the {\it NuSTAR} absorption line. Out of the $S$=10000 simulations (both FPMA and B spectra were simultaneously considered), we retrieved $N$=94 spectra in which a $\Delta\chi^2\leq-9$ was found. This leads to a statistical significance (1-94/10000), corresponding to $\simeq2.6\sigma$ (99.06\% confidence level). 

When we estimated the statistical significance of the two UFOs (three absorption lines) detected in the 2003 XMM spectrum, we followed the same technique discussed in \citet{lkg15} for IRAS 17020+4544. The detection confidence level for the two lines above 9 keV (UFO 2) is calculated taking into account the absorption line at 8.6 keV (UFO 1), i.e. the input model for the simulations is built adding the UFO 1 component to the continuum. The two statistically significant XMM absorption lines at 8.61 and 9.57 keV are therefore detected with a significance 2.7$\sigma$ and 3.6$\sigma$ (99.35\% and 99.97\% confidence levels), respectively.

\subsection{{\sc cloudy} modeling and line identification}
NGC 2992 is one of the handful of AGN in which a blueshifted absorption feature is detected with {\it NuSTAR}. In this Section we discuss the 2015 spectra and the 2003 EPIC pn data, which are both in a medium-high flux state (i.e. $5.8$ and $9.5\times 10^{-11}$ erg cm$^{-2}$ s$^{-1}$), using {\sc cloudy} generated tables. We took into account a self-consistent absorption model to estimate the column density and the ionization state of the UFOs.  We produced a grid model for \textsc{xspec} using \textsc{cloudy} 17, assuming a plane parallel geometry, with the flux of photons striking the illuminated face of the cloud given in terms of ionization parameter $U$ \citep{of06}; incident continuum modeled as in \citet{korista97}; constant electron density $\mathrm{n_e}=10^5$ cm$^{-3}$ and turbulence velocities in the range $v_{\rm turb}=500-2000$ km/s. Elemental abundances as in Table 7.1 of \textsc{cloudy} documentation\footnote{Hazy 1 version 17, p. 66: \url{http://viewvc.nublado.org/index.cgi/tags/release/c17.00/docs/hazy1.pdf?revision=11711&root=cloudy}}; grid parameters are $\log U=[1.00:5.00]$, $\log N_\mathrm{H}=[22.0:24.0]$.  

We start by fitting the high flux 2003 pn spectrum. When the primary continuum is convolved with one {\sc CLOUDY} component (zone 2) the fit improves ($\Delta\chi^2$=-38/3 d.o.f.) and the inclusion of a second absorber (zone 1: $\Delta\chi^2$=-18/3 d.o.f.) leads to a best fit $\chi^2$/dof=147/153=0.97: no strong residuals are seen throughout the 0.5-10 keV band (Fig. \ref{xmm03}, bottom panel) and best fit values can be found in Table \ref{bestfitwind}. When we leave the turbulence velocity as a variable parameter, no statistical improvement is found and the best fit value is $v_{\rm turb}>650$ km/s, with the parameter pegging at 2000 km/s. The outflowing velocity is a free parameter in our {\sc cloudy} components and two distinct velocities are required, $v_1=0.215\pm0.005c$ and $v_2=0.305\pm0.005c$.

 From a visual inspection of the best fitting model, the two absorption lines at 9.27 and 9.57 keV can be identified as Fe {\sc xxv} He-$\alpha$ and Fe {\sc xxvi} Ly-$\alpha$, with outflowing velocities v$_{\rm out}=0.31\pm0.01c$ and v$_{\rm out}=0.307\pm0.006c$, respectively. The third absorption line at 8.61 keV corresponds to Fe {\sc xxvi} Ly-$\alpha$, with an outflowing velocity v$_{\rm out}=0.209\pm0.006c$ and only an upper limit EW$<$40 eV is retrieved for the associated Fe {\sc xxv} He-$\alpha$ component. Fixing the outflow velocity to the one corresponding to Fe {\sc xxv} He-$\alpha$ to the {\sc cloudy} component, a statistically worse fit is obtained ($\Delta\chi^2$=+19/1 d.o.f.), clearly ruling out this solution. 
We also show in Table \ref{bestfitwind} the measured v$_{\rm out}/c$ values, using this identification. 
 We interpret the two absorption lines above 9 keV as Fe {\sc xxv} He-$\alpha$ and Fe {\sc xxvi} Ly-$\alpha$ with same outflowing velocity, but we cannot statistically rule out a fit in which they are modeled in terms of a single broad absorption line, leading to physical scenario with a single outflowing component (still at v$_{\rm out}\sim0.3c$) with a turbulence velocity $v_{\rm turb}\sim$10000 km/s.

\begin{figure}
  \epsfig{file=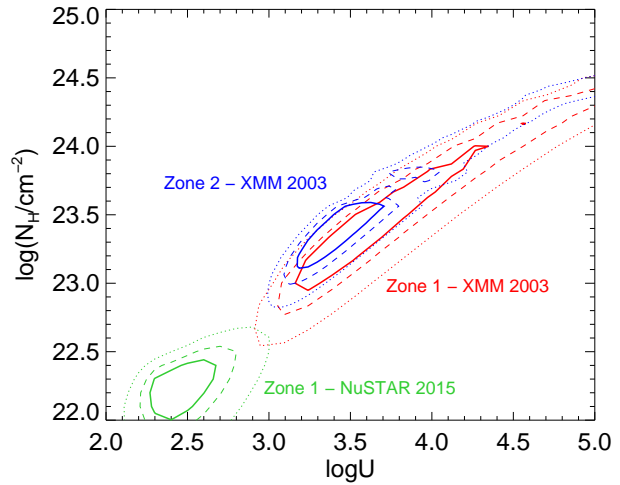, width=\columnwidth}
  \caption{Contour plots between column density and ionization parameter of the two outflowing components for the FPMA/B 2015 (in green) and EPIC pn 2003 spectra (red and blue). Solid, dashed and dotted lines indicate 68\%, 90\% and 99\% confidence levels.}
  \label{cont_U_nh}
\end{figure}

We then convolved the CLOUDY component (fixing the turbulence velocity to v$_{\rm turb}=2000$ km/s) with the model described in Sect. 3.2, removing the absorption line at 8.26 keV, to reproduce the 2015 {\it NuSTAR}-{\it Swift} data set. We find a very good fit ($\chi^2$/dof=375/415=0.91), with no statistically significant differences with other parameters previously inferred. We find a velocity $v_{\rm out}=0.21\pm0.01c$ for the outflowing gas, a column density $\log$(N$_{\rm H}/{\rm cm^{-2}})=22.25\pm0.25$ and a $\log U=2.45\pm0.25$. In this case, the absorption line is identified as Fe {\sc xxv} He-$\alpha$ and the measured $8.26^{+0.09}_{-0.12}$ keV centroid energy (Table \ref{bestfitwind}) can be converted in an outflowing velocity v$_{\rm out}=0.21\pm0.01c$. If we fix the outflow velocity to the one corresponding to Fe {\sc xxvi} Ly-$\alpha$ to the {\sc cloudy} component we obtain a worse fit ($\Delta\chi^2$=+13/1 d.o.f.).
The contour plots between the absorbing column density and the ionization parameter can be found in Fig. \ref{cont_U_nh}. One absorbing layer (zone 1: v$_{\rm out}\sim 0.21c$) is present in both XMM and {\it NuSTAR} data sets, while the other (zone 2: v$_{\rm out}\sim 0.30c$) appears only in the very bright state in 2003. The inferred parameters (N$_{\rm H}$, U, $v_{\rm out}$) for the first absorbing zone perfectly fall within the observed range of Ultra Fast Outflows detected with XMM \citep{tcr10,tcr11, tcr12} and Suzaku \citep{grt13,grm15}, and are different from the ones found for standard warm absorbers \citep{tcr13,lgd14}. The second UFO (zone 2: v$_{\rm out}\sim 0.30c$) is at the upper end of the outflow velocity distribution \citep[$\langle v_{\rm out}\rangle\sim0.1c$:][]{tcr10} and it is one of the fastest winds detected in a Seyfert galaxy, at modest accretion rate (L$_{\rm bol}$/L$_{\rm edd}\simeq$ 2--4\%). Such velocities are more typically seen around L$_{\rm bol}$/L$_{\rm edd}\sim1$ in high luminosity objects \citep{cbg02,rob09,tmv15,rbn18} or in NLS1s \citep{lkg15,pab17,ppf17, prm18}.

\section{Discussion}
\subsection{Origin of the Iron K$\alpha$ line}
\begin{figure}
 \epsfig{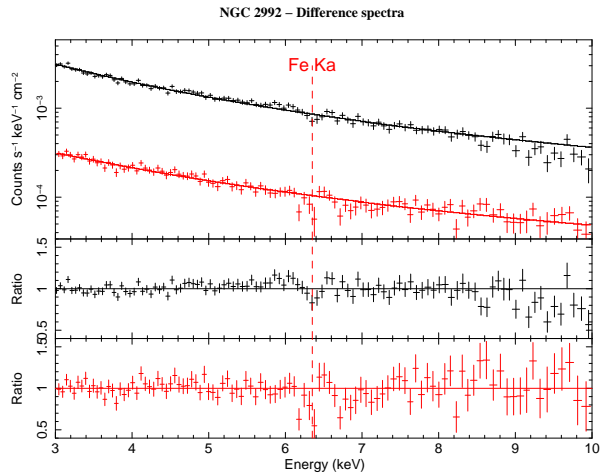}
  \caption{Difference spectra of the 2003 and 2010-11-28 observations are shown in black while 2010-05-26 (Obs. 3) and 2010-11-28 in red, respectively. Middle and bottom panels contain residuals when an absorbed power law model is applied to the data sets. Vertical dashed line indicates the rest-frame energy of the neutral Fe K$\alpha$ emission line.}
  \label{diffspec}
\end{figure}

XMM Observation no. 7 is one of the historical flux minima of the source and the measured EW of the Iron K$\alpha$ line (EW$_7$=$570\pm60$ eV) suggests that the primary continuum is very weak, allowing the reflection component to be clearly seen, with a very large reflection fraction ($R=4.50\pm1.10$). We therefore use this spectrum as input background for the 
2003 and 2010-05-26 (Obs. no. 3) XMM observations to construct difference spectra. Results in the 3-10 keV band are shown in Fig. \ref{diffspec} (top panel, in black and red, respectively) and when fitted with an absorbed power law the measured values are N$^{03}_{\rm H}=0.9\pm0.2\times10^{22}$ cm$^{-2}$, $\Gamma^{03}=1.90\pm0.05$ and N$^{10}_{\rm H}=1.2\pm0.3\times10^{22}$ cm$^{-2}$, $\Gamma^{10}=1.7\pm0.1$. This confirms that the main variable component is the nuclear emission and that it cannot be ascribed to column density changes along the line of sight. The different photon indices indicate that the spectrum is steeper when the source is brighter, as already discussed in \citet{shem06, rye09,upm16}. Residuals to this model are shown in middle and bottom panels of Fig. \ref{diffspec} and while strong residuals on the red wing of the Fe K$\alpha$ emission line can be clearly seen in the black data set, they disappear in the 2010 Obs. 3, being much more symmetric and centered around $\sim$6.6 keV. Even if we cannot totally discard systematic effects due to the EPIC pn CTI miscalibration we confirm that no positive residuals on the red wing of the Fe line due to relativistic effects are identified. \\
Summarizing the spectral analysis reported in Sect. 3, the Iron K$\alpha$ emission line complex in this object is likely the sum of three distinct components:\\
 - a narrow one due to reflection from cold, distant material (likely the molecular torus) which is constant, with a 2-10 keV flux F=$1.25\pm0.08\times 10^{-12}$ erg cm$^{-2}$ s$^{-1}$ and not responding to variations of the primary continuum;\\
- an unresolved, variable one which is more intense in brighter observations (Obs. 3, 9 and NuSTAR 2015). The emitting gas could be located in the outer parts of the accretion disk or in the BLR;\\
- a relativistic line emitted in the innermost regions of the accretion disk, which has been detected only in the 2003 XMM observation.

The optical emission of this source has been also widely investigated in the past \citep{veron80, allen99, trippe08} and \citet{gilli00} interpreted it in terms of a strongly variable underlying continuum with emission lines which are variable in flux but not in width. Indeed, the broad H$\alpha$ FWHM ranges from 2620 km/s in 1994 (near the X-ray minimum) to 2190 km/s in 1999 (1.5 months after the BeppoSAX high flux observation). If such material also emitted in the X-rays it would produce an Fe K$\alpha$ emission line with a line width $\sigma\sim20$ eV, which was not detected in the 2010 low flux {\it Chandra} HETG spectra \citep{mn17}. This suggests that the illuminated material (the outskirts of the accretion disk or the BLR) responds to variations of amplitude of the nuclear flux and the line intensity is hence proportional to the irradiating flux, similarly to what already found for NGC 2110 \citep{mmb15}.
\subsection{Energetics of the outflowing winds}
The black hole mass of NGC 2992 is estimated to be $5.2\times10^7$ M$_{\odot}$  via stellar velocity dispersion measurements \citep{wu02}, which leads to a relative fraction of the Eddington luminosity in the range L$_{\rm bol}$/L$_{\rm edd}\simeq$ (0.1-4)\%, using the bolometric correction from \citep{mar04}. 

The highest flux states discussed in this paper are the ones observed by {\it NuSTAR} in 2015 and by XMM-{\it Newton} in 2003, with 2-10 keV unabsorbed luminosities of 7.6$\times10^{42}$, 1.3$\times10^{43}$ erg/s, with bolometric luminosities L$_{\rm bol}$=$1.2\times10^{44}$, 2.4$\times10^{44}$ erg/s, corresponding to accretion rates L$_{\rm bol}$/L$_{\rm edd}$ of 2\% and 4\%, respectively. The spectral analysis discussed in Sect. 3.3 showed the presence of strongly blueshifted absorption features only in the 2015 {\it NuSTAR} and 2003 XMM spectra, i.e. when the accretion rate is larger than L$_{\rm bol}$/L$_{\rm edd}>2\%$. If modeled with a gas photoionised by the nuclear continuum along the line of sight, we find two accretion disk winds at different velocities  ($v_1\sim0.2c$ and $v_2\sim0.3c$). While the former is in the typical range inferred for UFOs in local Seyfert galaxies \citep{tcr12,grm15}, the latter is one of the fastest ever detected in a Seyfert galaxy at this low accretion rates. 

If we assume that the measured outflowing velocity corresponds to the escape velocity we can estimate the minimum radius $r_{\rm min}$ at which the accretion disk wind occurs. Following the parametrization on the geometry (conical) and on the density described in \citet{kne07} we can estimate the mass outflow rate $\dot{M}_{\rm out}$ and the kinetic power $\dot{E}_{\rm K}$ \citep[equations 3 and 4:][]{tcr13}.

For the first wind component (zone 1), found in both the 2015 and 2003 data sets, the outflowing velocity is $v_1\simeq65000$ km/s and it corresponds to an emission radius r$_{\rm min}=3.5\times10^{14}$ cm ($1.2\times10^{-4}$ pc, or $\sim25\ r_{\rm s}$ for a black hole mass M$_{\rm bh}$=$5.2\times10^7$ M$_{\odot}$). The lower limits for the mass outflow rate, total kinetic energy rate and momentum rate are reported in Table \ref{bestfitwind}. Assuming the more conservative value for the column density derived for the {\it NuSTAR} spectra and dividing the inferred $\dot{E}_{\rm K}$ by the Eddington luminosity (L$_{\rm Edd}=6.5\times10^{45}$ erg/s) we obtain log($\dot{E}_{\rm K}$/L$_{\rm Edd})=-2.9$, which locates this accretion disk wind in the UFO populated region in diagram 2 (left panel) in  \citet{tcr13}. We find that the total mechanical power $\dot{E}_{\rm K}\simeq$0.05L$_{\rm bol}$ is sufficient to switch on feedback mechanisms between the AGN and the host galaxy \citep[which can be as low as $\sim$0.005 L$_{\rm bol}$:][]{dsh05, he10}. We note, however, that if such winds are only triggered at the highest luminosities, their duty cycle is very small, and so will be the feedback effect \citep{ffs17}. This value is in perfect agreement with measurements in local sources \citep{tcr13,grm15} and much lower than the ones recently found for distant quasars, such as HS 0810+2554 \citep[z=1.51, $\dot{E}_{\rm K}\approx$9L$_{\rm bol}$:][]{cch16} and  MG J0414+0534 \citep[z=2.64, $\dot{E}_{\rm K}\approx$2.5L$_{\rm bol}$:][]{dvc18}. The total outflow momentum rate can be directly estimated as  $\dot{p}_{\rm out}=\dot{M}_{\rm out}v_{\rm out}$ and, if we compare it to the radiation momentum $\dot{p}_{\rm rad}$=$L_{\rm bol}/c$, we find $\dot{p}_{\rm out}\simeq0.5\ \dot{p}_{\rm rad}$. This suggest that the wind can be efficiently accelerated in terms of radiation pressure \citep[through Compton scattering;][]{kp03,ki10,ki10b}. 

When we take into account the second wind component (zone 2), only found in the 2003 XMM spectrum, the outflowing velocity is $v_2\simeq90000$ km/s and it corresponds to an emission radius r$_{\rm min}=1.7\times10^{14}$ cm ($6\times10^{-5}$ pc, or $\sim10\ r_{\rm s}$). The total outflow momentum rate from the two wind components is therefore $\dot{p}_{\rm out}\simeq6\ \dot{p}_{\rm rad}$ possibly indicating an additional/alternative accelerating mechanism, likely due to magnetic torques acting on the wind \citep{omm09,kfb12,fks17}

In NGC 2992 the fast wind features are sporadic, only appearing in the two brightest observations, at accretion rates larger than $\sim2\%$. The winds are launched in the innermost regions of the accretion disk (assuming that the outflowing velocity corresponds to the escape velocity), at a few tens of gravitational radii. We speculate that the wind can be more effectively accelerated when the luminosity and thus the accretion rate is higher, somehow reaveling the innermost regions of the disk. Indeed, such accretion disk winds are accompanied by a broad component of the Iron K$\alpha$ emission line.
This case is clearly different from other highly accreting objects, such as PDS 456 \citep{mrn17,mrb17}, where a correlation between the outflowing velocity and the luminosity is observed. Indeed, a response of the relativistic outflowing gas to the variations of the nuclear continuum (in particular the ionization stage) has been recently shown for PDS 456 and IRAS 13224-3809 \citep{mrn16,pab17,ppf17, prm18}. However, while these two sources are accreting close to L$_{\rm Edd}$, NGC 2992 is accreting only at a few per cents of the Eddington luminosity. 
\section{Conclusions}
NGC 2992 has been observed with all major X-ray satellites since 1978 showing a strong variability (up to a factor 20) on timescales of weeks/months. 
We presented a broadband spectroscopic analysis of eight XMM observations performed in 2010 and one in 2013, of the 2015 simultaneous {\it Swift} and {\it NuSTAR} data and of an additional 2003 XMM observation, once corrected for pile-up effects. The source was always in a faint state in 2010/2013 but {\it NuSTAR} showed an increase in the 2-10  keV flux up to 6$\times10^{-11}$ erg cm$^{-2}$ s$^{-1}$ in 2015. Possible evidence of an Ultra Fast Outflow with velocity $v_1=0.21\pm0.01c$ (at 2.6$\sigma$ confidence level) is found in this {\it NuSTAR} observation. This spectral component is not present in the 2010/2013 low flux spectra, but it is detected in the bright 2003 XMM observation (at 2.7$\sigma$ confidence level).
Using CLOUDY generated tables, the two zones of absorption can be modeled in terms of a highly ionized gas with column densities in the range N$_{\rm H}=10^{22}$--$10^{23}$ cm$^{-2}$ and ionization parameters $\log U=2.5$--3.5. We find that the total kinetic energy rate of the {\it NuSTAR} outflow is $\approx5\%$ L$_{\rm bol}$, sufficient to switch on feedback mechanisms on the host galaxy. Such accretion disk winds are sporadic in NGC 2992, they arise from a few tens of gravitational radii and have been detected only when the accretion rate of the source exceeds 2\% of the Eddington luminosity. This could be indicative of an accretion disk which reveals its innermost regions when the flux rises. In the future, further X-ray observations will test this scenario in more detail.

The analysis of the low flux 2010/2013 XMM data allowed us to determine that the Iron K$\alpha$ emission line complex in this object is likely the sum of three distinct components. The first one is narrow and due to reflection from cold, distant material (likely the molecular torus) not responding to variations of the primary continuum and with a constant 2--10 keV flux F=$1.25\pm0.08\times 10^{-12}$ erg cm$^{-2}$ s$^{-1}$. The second one is an unresolved, variable component which is more intense in brighter observations (Obs. 3 and 9 of the 2010/2013 monitoring campaign). The last one is broad (EW$_{\rm br}=230^{+60}_{-40}$ eV) and due to relativistic effects in the innermost regions of the accretion disk, only detected in the 2003 XMM observation.

\section*{ACKNOWLEDGEMENTS}
We thank the anonymous referee for her/his comments and suggestions which improved the manuscript. AM, SB and GM acknowledge financial support from the European Union Seventh Framework Programme (FP7/2007-2013) under grant agreement n.312789. AM, GM acknowledge financial support from the Italian Space Agency under grant ASI/INAF I/037/12/0-011/13. SB acknowledges financial support from the Italian Space Agency under grant ASI-INAF I/037/12/0. EN acknowledges funding from the European Union's Horizon 2020 research and innovation programme under the Marie Sk\l{l}odowska-Curie grant agreement No. 664931. 

\section*{Appendix A}
In this Appendix, we discuss instrumental effects induced by pile-up in the 2003 XMM observation. NGC 2992 was targeted by XMM in Full Frame mode in 2003, for a total elapsed time of 27 ks. With an average 0.5-10 keV count rate of 19.32$\pm$0.39 cts/s, the observation was inevitably affected by pile-up. We reduced the data set following the same approach described in Sect. 2.1 (for calculating source extraction radius, optimal time cuts for flaring particle background and energy binning) without pile-up corrections. This spectrum (not excised) was then compared to the one used for our analysis, in which pile-up is removed. In principle, there is no reason to expect that pile-up could create narrow line-like spurious features, it produces a flatter spectral shape, due to the false detection of soft X-ray photons at higher energies. We therefore fitted the two spectra with a phenomenological model composed of a power law plus a narrow Gaussian line at 6.4 keV, in the 3-10 keV band. As expected, we obtain a harder photon index for the piled up spectrum ($\Gamma_1=1.50\pm0.01$), compared to the one in which such an instrumental effect is removed ($\Gamma_2=1.72\pm0.02$). The high energy UFO features in NGC 2992 are completely smeared out in the piled up spectrum and also part of the broad component of the Fe K$\alpha$ is hidden, as already discussed in \citet{mdb10}.
They appear in the pile-up corrected spectrum and, even if we exclude the central region of the source up to 15 arcsec \citep[instead of 10 arcsec, as in][]{sym10}, we find the same slope for the continuum compared to the one reported in Sect. 3.3 and the same absorption features, with larger uncertainties. 
\bibliographystyle{mn2e}
\bibliographystyle{mn2e}
\bibliography{sbs} 

\end{document}